\documentclass[preprint,aps,pre]{revtex4}

\usepackage{epsfig}
\usepackage{psfig}

\usepackage{doublespace}

\newcommand{\erf}{{\rm erf}}
\newcommand{\bo}[1]{\mbox{\boldmath \rm $#1$}}

\begin{document}
\title{Replica symmetric evaluation of the information transfer in a
two-layer network in presence of continuous and discrete stimuli}

\author{Valeria Del Prete\thanks{delprete@sissa.it}  and Alessandro Treves}
 
\affiliation{SISSA, Programme in Neuroscience, via Beirut 4, Trieste, Italy}

\bibliographystyle{unsrt}
\begin{abstract}
In a previous report we have evaluated analytically the mutual information between the 
firing rates of $N$ independent units and a set of multi-dimensional continuous+discrete
stimuli, for a finite population size and in the limit of large noise. Here, we extend 
the analysis to the case of two interconnected populations, where input units activate 
output ones via gaussian weights and a threshold linear transfer function.
We evaluate the information carried by a population of $M$ output units, again about 
continuous+discrete correlates. The mutual information is evaluated solving saddle point
equations under the assumption of replica symmetry, a method which, by taking into 
account only the term linear in $N$ of the input information, is equivalent to assuming 
the noise to be large. Within this limitation, we analyze the dependence of the 
information on the ratio $M/N$, on the selectivity of the input units and on the level 
of the output noise. We show analytically, and confirm numerically, that in the limit 
of a linear transfer function and of a small ratio between output and input noise, the 
output information approaches asymptotically the information carried in input. Finally, 
we show that the information loss in output does not depend much on the structure of the
stimulus, whether purely continuous, purely discrete or mixed, but only on the position 
of the threshold nonlinearity, and on the ratio between input and output noise.
\end{abstract}

 
\maketitle 

\section{Introduction}
Recent analyses of extracellular recordings performed in two motor areas of
behaving monkeys have tried to clarify how information about movements is
trasmitted and received from higher to lower stages of processing, and
to identify distinct roles of the two areas in the planning and execution of
movements \cite{vale+01c}. Although this study failed to produce clearcut results,
it remains interesting to try and understand, from a more theoretical point of view, 
how information about multi-dimensional correlates of neural activity may be 
transmitted from the input to the output of a simple network. In fact, a theoretical 
study is still lacking, which explores how the coding of stimuli with continuous 
as well as discrete dimensions is transferred across a network. 

Information theory \cite{Sha+48} has been widely used in the theory of  
communication,  
in presence of both binary \cite{NP1,KNP,TKP} and linear \cite{Guid,NP}
or weakly non-linear \cite{vale+01a} channels. 
Moreover it has been recently proposed as an effective tool to explore 
the coding properties of neurons (see for example \cite{Rol+98,Bia+91,Rie+96}), via 
both direct estimates from real data (for a review see \cite{proc30})
and pure theoretical modelling \cite{Tre+92,Tre+95,Tre+98,pap35,sompo+00,vale+01b}.  

The {\it mutual information} provides a quantitative 
and flexible measure of the efficiency 
of single cells or of population of cells in coding external stimuli and events 
relevant to behaviour: high values of the mutual information are obtained when 
the correlates can be discriminated with a small uncertainty on the basis of the 
neural responses; moreover the same formalism can be adapted 
to explore different types of code, from simple time-averaged rates, to more 
sophisticated descriptions, 
where the exact temporal sequence of action potentials is considered to be relevant. 

In a previous report \cite{vale+01b} the mutual information between the time averaged rates 
of a finite population of $N$ units and a set of correlates, 
which have both a discrete and a continuous angular dimension, has been evaluated
analytically in the limit of large noise. This parameterization of the correlates
can be applied to movements performed in a given direction and classified according to
different "types"; yet it is equally applicable to other correlates, like visual stimuli 
characterized by an orientation and a discrete feature (colour, shape, etc..), or in 
general to any correlate which can be identified by an angle and a "type".
In this study, we extend the analysis performed for one population, to consider
two interconnected areas, and we evaluate the mutual information between 
the firing rates of a finite population of $M$ output neurons and a set of
continuous+discrete stimuli, given that the rate distribution in input is known.
In input, a threshold nonlinearity has been shown to lower the information about the 
stimuli in a simple manner, which can be expressed as a renormalization of the noise 
\cite{vale+01b}. How does the information in the output depend on the same nonlinearity?
How does it depend on the noise in the output units? Is the information transmission from input 
to output sensitive 
to the structure of the correlate, whether discrete or continuous? 

We address these issues by calculating the mutual information, using the replica trick 
and under the assumption of replica symmetry (see for example \cite{meza+87}).

Saddle point equations are solved numerically. We analyze how the information 
trasmission depends on the parameters of the model, i.e. the level of output and 
input noise, on the ratio between the two population sizes, as well as on the tuning 
curve with respect to the continuous correlate, and on number of discrete correlates. 

The input-output transfer function is a crucial element in the model. Many earlier theoretical and simulation studies 
\cite{Ami+89} have mainly focused on binary and 
the sigmoidal functions; yet more recent investigations \cite{Tre+90,Tre+98b} have shown 
that the current-to-frequency transduction typical of real neurons
is well captured,
away from saturation, by a threshold-linear function.
Such a function combines the threshold of real neurons, the linear behaviour
typical of pyramidal neurons above threshold, and the accessibility to a full
analytical treatment \cite{Tre+90b,Tre+95}, as demonstrated here, too. For the 
sake of analytical feasibility, however, we take the input units to be purely
gaussian. Therefore it should be kept in mind, in considering the final results,
that the threshold nonlinearity is only applied to the output units.

\section{The model}
In analogy to the model studied in \cite{vale+01b} we consider a set of $N$ input units 
which fire to an external
continuous+discrete stimulus, parameterized by an angle $\vartheta$ and a discrete 
variable $s$, with a gaussian 
distribution:
\begin{equation}
p(\{\eta_j\}|\vartheta,s)=\prod_{j=1}^N \frac{1}{\sqrt{2\pi\sigma^2}}
exp-\left[\left(\eta_j-\tilde{\eta}_j(\vartheta,s)
\right)^2/2\sigma^2\right];
\label{dist}
\end{equation}
$\eta_j$ is the firing rate in one trial of the $j^{th}$ input neuron, while the mean 
of the distribution, $\tilde{\eta}_j(\vartheta,s)$ is written:
\begin{equation}
\tilde{\eta}_j(\vartheta,s)=\varepsilon^j_s\bar{\eta}_j(\vartheta)+(1-\varepsilon_s^j)\eta^f;
\label{tuning_tot}
\end{equation}
\begin{equation}
\bar{\eta}_j(\vartheta-\vartheta^0_j)=
\eta^0
\cos^{2m}\left(\frac{\vartheta-\vartheta^0_j}{2}\right);
\label{tuning2}
\end{equation}
where $\varepsilon_s^j$ is a quenched random variable distributed between $0$
and $1$, $\vartheta_i^0$ is the preferred direction for neuron $i$.
According to eq.(\ref{tuning_tot}) neurons fire at an average firing rate which 
modulates with $\vartheta$ with amplitude $\varepsilon_s$, or takes a fixed value 
$\eta^f$, independently of $\vartheta$, with amplitude $1-\varepsilon_s$.

We assume that quenched disorder is uncorrelated and identically distributed
across units and across the $K$ discrete correlates, and that for each neuron 
all preferred directions are equally likely:
\begin{equation}
\varrho(\{\varepsilon^i_s\})=\prod_{i,s}\varrho(\varepsilon_s^i)=\left[\varrho(\varepsilon)\right]^{NK}
\label{ro_eps}
\end{equation}
$$
\varrho(\{\vartheta^0_i\})=\left[\varrho(\vartheta^0)\right]^N=\frac{1}{(2\pi)^N}.
$$

In \cite{vale+01b} it has been shown that a cosinusoidal shaped function as in 
eq.(\ref{tuning_tot}) 
is able to capture the main features of directional tuning of real neurons in motor cortex.   
Moreover it has been shown that the presence of negative firing rates in the distribution (\ref{dist}), 
which is not biologically plausible, does not alter information values, with respect to a more
realistic choice for the firing distribution, in that it leads to the same curves
except for a renormalization of the noise.

Output neurons are connected to input neurons via uncorrelated gaussian connection weights 
$J_{ij}$. Each output neuron performs a linear summation of the inputs; the outcome 
is distorted by a 
gaussian distributed fast noise $\delta_i$ and then thresholded, as in the following:  
\begin{equation}
\xi_i=[\xi_i^0+\sum_j c_{ij} J_{ij} \eta_j
+\delta_i]^+;\,\,\,\,i=1..M,j=1..N
\label{output} 
\end{equation}

In eq.(\ref{output})$\xi_i^0$ is a threshold term,
$c_{ij}$ is a (0,1) binary variable, with mean $c$, which expresses the sparsity or 
dilution of
the connectivity matrix, and
\begin{equation} 
\langle (J_{ij})^2\rangle=\sigma^2_J;\,\,\,\,\langle J_{ij}\rangle=0;
\end{equation}
\begin{equation} 
\langle (\delta_{i})^2\rangle=\sigma^2_\delta;\,\,\,\,\langle \delta_{i}\rangle=0;
\end{equation}
\begin{eqnarray}
p(c_{ij}\!=\!1)&=&c;\nonumber\\
p(c_{ij}\!=\!0)&=&1-c;
\end{eqnarray}
\begin{equation}
[x]^+=x\Theta(x).
\end{equation}

\section{Analytical estimation of the mutual information}

We aim at estimating the mutual information between the output patterns of
activity and the continuous+discrete stimuli:
\begin{equation}
I(\{\xi_i\},\vartheta\otimes s)=\left\langle\sum_{s=1}^K\int \!d\vartheta\!
\int \prod_i d\xi_ip(\vartheta,s)
p(\{\xi_i\}|\vartheta,s)\log_2\frac{p(\{\xi_i\}|\vartheta,s)}{p(\{\xi_i\})}
\right\rangle_{\varepsilon,\vartheta^0,c,J,\delta}; 
\label{info} 
\end{equation}
\begin{equation}
p(\{\xi_i\}|\vartheta,s)=\int \prod_j d\eta_j
p(\{\xi_i\}|\{\eta_j\})p(\{\eta_j\}|\vartheta,s);
\label{xi_eta}
\end{equation}
where the distribution $p(\{\xi_i\}|\{\eta_j\})$ is
determined by the threshold linear relationship
(\ref{output}), $p(\{\eta_j\}|\vartheta,s)$ is given in eq.(\ref{dist}) and
$\langle .. \rangle_{\varepsilon,\vartheta^0,c,J,\delta}$ is a short notation
for the average across the quenched variables
$\{\varepsilon_s^i\}$,$\{\vartheta_i^0\}$,$\{J_{ij}\}$,$\{c_{ij}\}$ and on
the fast noise $\{\delta_i\}$.

Contrary to the other quenched variables, 
$\{\varepsilon_s^i\}$,$\{\vartheta_i^0\}$,$\{J_{ij}\}$,$\{c_{ij}\}$, the variable 
$\delta_i$ in eq.(\protect\ref{output}) is annealed:
integration of relationships (\protect\ref{output}) across a zero
mean gaussian distribution of $\delta_i$ with variance $\sigma^2_\delta$
yields a gaussian distribution in $\xi_i$ with variance $\sigma^2_\delta$.
We assume that the stimuli are equally likely: $p(\vartheta,s)=1/2\pi K$.

Eq.(\ref{info}) can be written as:
\begin{equation}
I(\{\xi_i\},\vartheta\otimes s)=\left\langle
H(\{\xi_i\}|\vartheta,s) \right\rangle_{\vartheta,s}-H(\{\xi_i\})
\end{equation}
with:
\begin{equation}
\left\langle H(\{\xi_i\}|\vartheta,s)\right\rangle_{\vartheta,s}=
\left\langle\sum_{s}\int \!d\vartheta\! \int \prod_i d\xi_i 
p(\vartheta,s) p(\{\xi_i\}|\vartheta,s)\log_2
p(\{\xi_i\}|\vartheta,s)
\right\rangle_{\varepsilon,\vartheta^0,c,J,\delta};
\label{equiv}
\end{equation}
\begin{eqnarray}
&&H(\{\xi_i\})=\left\langle\sum_{s}\int \!d\vartheta\! \int \prod_i d\xi_i 
p(\vartheta,s) p(\{\xi_i\}|\vartheta,s)\right.\nonumber\\
&&\left.\log_2\left[
\sum_{s^\prime}\int \!d\vartheta^\prime p(s^\prime,\vartheta^\prime)
p(\{\xi_i\}|\vartheta^\prime,s^\prime)\right]
\right\rangle_{\varepsilon,\vartheta^0,c,J,\delta}.
\label{outent}
\end{eqnarray}

The analytical evaluation of $\left\langle
H(\{\xi_i\}|\vartheta,s)\right\rangle_{\vartheta,s}$ can be performed
inserting eq.(\ref{xi_eta}) in the expression (\ref{equiv}), and using the
replica trick \cite{meza+87} to get rid of the sums under logarithm; 
 since these sums already 
multiply the logarithm, all replica indexes run from $1$ up to $n+1$:
\begin{eqnarray}
&&\left\langle H(\{\xi_i\}|\vartheta,s)\right\rangle_{\vartheta,s}=\nonumber\\
&&\lim_{n\rightarrow 0}\frac{1}{n \ln2}\left (\left\langle\sum_s\int d\vartheta
p(\vartheta,s) \int \prod_{j,\alpha} d\eta_j^\alpha
\prod_{j,\alpha}p(\eta_j^\alpha|\vartheta,s)\int \prod_i d\xi_i
\prod_{j,\alpha}
p(\xi_i|\{\eta_j^{\alpha}\})\right\rangle_{\varepsilon,\vartheta^0,c,J,\delta}\!\!\!\!-1\right).
\label{replica}
\end{eqnarray}

To take into account the threshold-linear relation (\ref{output}) we
consider the following equalities:
\begin{eqnarray}
&&\int d\xi_i \prod_\alpha p(\xi_i|\{\eta_j^\alpha\})=\prod_\alpha
p(\xi_i=0|\{\eta_i^\alpha\})+\int_0^\infty d\xi_i \prod_\alpha
p(\xi_i|\{\eta_i^\alpha\})=\nonumber\\
&&\int_{-\infty}^0\!\!\prod_\alpha d\xi_i^{\alpha}
\delta(\xi_i^\alpha\!-\!\xi_i^0-\sum_j c_{ij} J_{ij} \eta_j^\alpha
-\delta_i^\alpha)+\int_0^\infty\!\! d\xi_i\prod_\alpha
\delta(\xi_i\!-\!\xi_i^0-\sum_j c_{ij} J_{ij} \eta_j^\alpha -\delta_i^\alpha).
\label{anna}
\end{eqnarray}

Inserting eq.(\ref{anna}) in eq.(\ref{replica}) one obtains:
\begin{eqnarray}
&&\left\langle H(\{\xi_i\}|\vartheta,s)\right\rangle_{\vartheta,s}=
\lim_{n\rightarrow 0}\frac{1}{n \ln2}\left (\sum_s\int d\vartheta
p(\vartheta,s) \int \prod_{j,\alpha} d\eta_j^\alpha
\left\langle\prod_{j,\alpha}p(\eta_j^\alpha|\vartheta,s)\right\rangle_{\varepsilon,\vartheta^0}\right.\nonumber\\
&&\left.\prod_i \left[\int_{-\infty}^0 \prod_\alpha d\xi_i^\alpha 
\left\langle \prod_\alpha \delta(\xi_i^\alpha-\xi_i^0-\sum_j c_{ij} J_{ij}
\eta_j^\alpha
-\delta_i^\alpha)\right\rangle_{c,J,\delta}\right.\right.\nonumber\\
&&\left.\left.+\int_0^\infty
d\xi_i  \left\langle \prod_\alpha \delta(\xi_i-\xi_i^0-\sum_j c_{ij} J_{ij}
\eta_j^\alpha
-\delta_i^\alpha)\right\rangle_{c,J,\delta}\right] -1\right).
\label{replica2}
\end{eqnarray} 

The average across the quenched disorder $c$,$J$,$\delta$ in eq.(\ref{replica2}) can
be performed in a very similar way as shown in \cite{Tre+98b}: using the integral
representation for each $\delta$ function, gaussian integration across
$J$,$\delta$ is  standard; the average on $\{c_{ij}\}$ can be performed
assuming large the number $N$ of input neurons, that is for very small $c$. 

The final outcome for $\left\langle H(\{\xi_i\}|\vartheta,s)\right\rangle_{\vartheta,s}$ 
reads:
\begin{eqnarray}
&&\left\langle H(\{\xi_i\}|\vartheta,s)\right\rangle_{\vartheta,s}=
\lim_{n\rightarrow 0}\frac{1}{n \ln2}\left (\sum_s\int d\vartheta
p(\vartheta,s) \int \prod_{j,\alpha} d\eta_j^\alpha
\left\langle\prod_{j,\alpha}p(\eta_j^\alpha|\vartheta,s)\right\rangle_{\varepsilon,\vartheta^0}\right.\nonumber\\
&&\left.\left[\int_{-\infty}^0 
\prod_\alpha d\xi^\alpha\int \prod_\alpha \frac{dx^\alpha}{2\pi}
e^{-(\sigma^2_\delta/2)\sum_\alpha (x^\alpha)^2}
e^{-(C\sigma^2_J/2N) \sum_{\alpha,\beta}x^\alpha
x^\beta\sum_j\eta_j^\alpha\eta_j^\beta}
e^{-i\sum_\alpha(\xi^\alpha-\xi_0)x^\alpha}\right.\right.\nonumber\\
&&\left.\left.+\int^{\infty}_0   d\xi \int \prod_\alpha
\frac{dx^\alpha}{2\pi} e^{-(\sigma^2_\delta/2)\sum_\alpha
(x^\alpha)^2} e^{-(C\sigma^2_J/2N) \sum_{\alpha,\beta}x^\alpha
x^\beta\sum_j\eta_j^\alpha\eta_j^\beta} e^{-i(\xi-\xi_0)\sum_\alpha
x^\alpha}\right]^M -1\right),  \label{replica3} \end{eqnarray}  where we have
put $c\rightarrow C/N$.

Integration on $\{x^\alpha\}$ is straightforward. Integration on $\{\eta_i^\alpha\}$ 
can be performed introducing $(n+1)^2$ auxiliary variables
$z_{\alpha\beta}=\frac{1}{N}\sum_j\eta_j^\alpha\eta_j^\beta$ via $\delta$
functions expressed in their integral representation. Considering the
expression (\ref{dist}) for the input distribution and with some rearrangement
of the terms the final result can be expressed as:
\begin{eqnarray}
&&\left\langle H(\{\xi_i\}|\vartheta,s)\right\rangle_{\vartheta,s}=\label{equiv2}\\
&&\lim_{n\rightarrow 0}\frac{1}{n \ln2}\left(\int\prod_{\alpha,\beta} 
\frac{dz_{\alpha\beta}}{2\pi/N}\int\prod_{\alpha,\beta} 
d\tilde{z}_{\alpha\beta} e^{iN\sum_{\alpha\beta}z_{\alpha\beta}
\tilde{z}_{\alpha\beta}}e^{-\frac{N}{2}Tr \ln \Sigma}
\sum_s\!\!\int\!\! d\vartheta p(\vartheta,s) \left\langle 
e^{-\sum_{\alpha,\beta}(\delta_{\alpha\beta}-\Sigma^{-1}_{\alpha\beta})
\tilde{\eta}(\vartheta,s)^2/2\sigma^2}\right\rangle_{\varepsilon,\vartheta^0}^N\right.\nonumber\\
&&\left.e^{-\frac{M}{2}Tr \ln G}
\left[\int_{-\infty}^0 
\prod_\alpha \frac{d\xi^\alpha}{\sqrt{2\pi}} e^{-\sum_{\alpha,\beta}(\xi^\alpha-\xi_0)
(G^{-1}_{\alpha\beta}/2)(\xi^\beta-\xi_0)}+\int^{\infty}_0 
 \frac{d\xi}{(2\pi)^{\frac{n+1}{2}}} e^{-\sum_{\alpha,\beta}
(G^{-1}_{\alpha\beta}/2)(\xi^-\xi_0)^2}\right]^M-1\right),
\nonumber
\end{eqnarray}
where:
\begin{eqnarray}
&&\Sigma_{\alpha\beta}=\delta_{\alpha\beta}+2\sigma^2 i
\tilde{z}_{\alpha\beta};\nonumber\\  
&&G_{\alpha\beta}=\sigma^2_\delta\delta_{\alpha\beta}+C\sigma^2_Jz_{\alpha\beta}.
\end{eqnarray}

The evaluation of 
$H(\{\xi_i\})$, eq.(\ref{outent}), can be carried out in a very similar way,
introducing replicas in the continuous+discrete stimulus space.
The final result reads:
\begin{eqnarray}
&&H(\{\xi_i\})=\lim_{n\rightarrow 0}\frac{1}{n \ln2}\left(\int\prod_{\alpha,\beta} 
\frac{dz_{\alpha\beta}}{2\pi/N}\int\prod_{\alpha,\beta} 
d\tilde{z}_{\alpha\beta} e^{iN\sum_{\alpha\beta}z_{\alpha\beta} 
\tilde{z}_{\alpha\beta}}e^{-\frac{1}{2}Tr \ln \Sigma}\right.\nonumber\\
&&\left. \sum_{s_1..s_{n+1}}\int
d\vartheta_1..d\vartheta_{n+1}\left[p(\vartheta,s)\right]^{n+1} \left\langle 
e^{-\sum_{\alpha,\beta}(\delta_{\alpha\beta}-\Sigma^{-1}_{\alpha\beta})
\tilde{\eta}(\vartheta_\alpha,s_\alpha)\tilde{\eta}(\vartheta_\beta,s_\beta)/2\sigma^2}
\right\rangle_{\varepsilon,\vartheta^0}^N\right.\label{outent2}\\
&&\left.e^{-\frac{M}{2}Tr \ln G} \left[\int_{-\infty}^0 
\prod_\alpha \frac{d\xi^\alpha}{\sqrt{2\pi}} e^{-\sum_{\alpha,\beta}(\xi^\alpha-\xi_0)
(G^{-1}_{\alpha\beta}/2)(\xi^\beta-\xi_0)}+\int^{\infty}_0 
 \frac{d\xi}{(2\pi)^{\frac{n+1}{2}}} e^{-\sum_{\alpha,\beta}
(G^{-1}_{\alpha\beta}/2)(\xi^-\xi_0)^2}\right]^M-1\right).
\nonumber
\end{eqnarray}

\section{Replica symmetric solution}

The integrals in eq.(\ref{equiv2}),(\ref{outent2}) cannot be solved without
resorting to an approximation. In analogy to what is used in
\cite{Tre+95,Tre+98b},  we use a saddle-point approximation (which in general
would be valid in the limit  $M,N\rightarrow\infty$) and we assume replica
symmetry \cite{meza+87} in the parameters $\{z_{\alpha\beta}\}$,
$\{\tilde{z}_{\alpha\beta}\}$. This allows to explicitely invert  and
diagonalize the matrices $G$,$\Sigma$: \begin{eqnarray}
z_{\alpha\alpha}&=&z_0(n);\,\,\,z_{\alpha\neq\beta}=z_1(n);\nonumber\\
i\tilde{z}_{\alpha\alpha}&=&\tilde{z}_0(n);\,\,\,i\tilde{z}_{\alpha\neq\beta}=-\tilde{z}_1(n);
\end{eqnarray}

In \cite{Tre+98b} it has been shown that the replica 
symmetric solution for the information transmitted by a threshold-linear net is stable 
in most of the phase diagram. A detailed study of the stability of the RS solution in the specific 
case of mixed continuous and discrete stimuli will be presented elsewhere \cite{Isaac+02}.
The saddle point approximation seems to have more subtle implications in the present 
situation, as it will be discussed in the next section and in the final discussion.  

In replica symmetry the mutual information can be expressed as follows:
\begin{eqnarray}
I(\{\xi_i\},\vartheta\otimes s)&=&\lim_{n\rightarrow 
0}\frac{1}{n\ln2}\left\{e^{N\left[(n+1)z_0^A\tilde{z}_0^A-n(n+1)z_1^A\tilde{z}_1^A-
\frac{r}{2}\left(Tr\ln
G(z_0^A,z_1^A)+F(z_0^A,z_1^A)\right)-\frac{1}{2}Tr\ln\Sigma(\tilde{z}_0^A,\tilde{z}_1^A)
-H^A(\tilde{z}_0^A,\tilde{z}_1^A)\right]}\right.\nonumber\\
&&\left.-e^{N\left[(n+1)z_0^B\tilde{z}_0^B-n(n+1)z_1^B\tilde{z}_1^B-
\frac{r}{2}\left(Tr\ln
G(z_0^B,z_1^B)+F(z_0^B,z_1^B)\right)-\frac{1}{2}Tr\ln\Sigma(\tilde{z}_0^B,\tilde{z}_1^B)-
H^B(\tilde{z}_0^B,\tilde{z}_1^B)\right]}\right\},
\label{info_lim}
\end{eqnarray}
with
\begin{equation}
F(z_0,z_1)\!=\!\!-2\ln\left[\int_{-\infty}^0 
\prod_\alpha \frac{d\xi^\alpha}{\sqrt{2\pi}} e^{-\sum_{\alpha,\beta}(\xi^\alpha-\xi_0)
(G^{-1}_{\alpha\beta}/2)(\xi^\beta-\xi_0)}\!+\!\int^{\infty}_0 
\!\! \frac{d\xi}{(2\pi)^{\frac{n+1}{2}}} e^{-\sum_{\alpha,\beta}
(G^{-1}_{\alpha\beta}/2)(\xi^-\xi_0)^2}\right];
\label{F}
\end{equation}
\begin{equation}
H^A(\tilde{z}_0,\tilde{z}_1)=-\frac{1}{N}\ln\left[\sum_s\int d\vartheta
p(\vartheta,s) \left\langle 
e^{-\sum_{\alpha,\beta}\left(\delta_{\alpha\beta}-\Sigma^{-1}_{\alpha\beta}\right)
\tilde{\eta}(\vartheta,s)^2/2\sigma^2}\right\rangle_{\varepsilon,\vartheta^0}^N\right];
\label{HA}
\end{equation}
\begin{equation}
H^B(\tilde{z}_0,\tilde{z}_1)\!=\!-\frac{1}{N}\ln\left[\sum_{s_1..s_{n+1}}\!\int\!
d\vartheta_1..d\vartheta_{n+1} [p(\vartheta,s)]^{n+1} \left\langle 
e^{-\sum_{\alpha,\beta}\left(\delta_{\alpha\beta}-\Sigma^{-1}_{\alpha\beta}\right) 
\tilde{\eta}(\vartheta_\alpha,s_\alpha)\tilde{\eta}(\vartheta_\beta,s_\beta)/2\sigma^2}
\right\rangle_{\varepsilon,\vartheta^0}^N\right].
\label{HB}
\end{equation}

We have set $r=\frac{M}{N}$ and
$z_0^{A,B}$,$\tilde{z}_0^{A,B}$,$z_1^{A,B}$,$\tilde{z}_1^{A,B}$ are the
solutions of the saddle point equations:
\begin{eqnarray}
z_0^{A,B}&=&\frac{\partial}{\partial\tilde{z}_0}\left[\frac{1}{2}
Tr\ln\Sigma(\tilde{z}_0,\tilde{z}_1)+H^{A,B}(\tilde{z}_0,\tilde{z}_1)\right];\nonumber\\
z_1^{A,B}&=&-\frac{1}{n}\frac{\partial}{\partial\tilde{z}_1}\left[\frac{1}{2}
Tr\ln\Sigma(\tilde{z}_0,\tilde{z}_1)+H^{A,B}(\tilde{z}_0,\tilde{z}_1)\right];\nonumber\\
\tilde{z}_0^{A,B}&=&\frac{\partial}{\partial z_0}\frac{r}{2}\left[
Tr\ln G(z_0,z_1)+F(z_0,z_1)\right];\nonumber\\
\tilde{z}_1^{A,B}&=&-\frac{1}{n}\frac{\partial}{\partial z_1}\frac{r}{2}\left[
Tr\ln G(z_0,z_1)+F(z_0,z_1)\right].
\end{eqnarray}

All the equations must be evaluated in the limit $n\rightarrow 0$.
It is easy to check that all terms in the exponent in eq.(\ref{info_lim}) are
order $n$. In fact, since when $n\rightarrow 0$ only one replica remains, one
has:
\begin{eqnarray}
\lim_{n\rightarrow 0} Tr\ln G(z_0,z_1)\!+\!F(z_0,z_1)&=&0\rightarrow \nonumber \\
Tr\ln G(z_0,z_1)+F(z_0,z_1)&\simeq &n \frac{\partial}{\partial n} [Tr\ln
G(z_0,z_1)+F(z_0,z_1)]_{|_{n=0}}.
\end{eqnarray}

Therefore, from the saddle point equations, $\tilde{z}_0^{A,B}$ are order $n$
and $Tr\ln\Sigma$ is also order $n$: 
\begin{equation}
Tr\ln\Sigma\simeq n \frac{\partial}{\partial n} Tr\ln \Sigma_{|_{n=0}}.
\end{equation}

Since $\tilde{z}_0^{A,B}=n\tilde{\tilde{z}}_0^{A,B}$, it is easy to check by
explicit evaluation that, when $n\rightarrow 0$, all the $n+1$ diagonal terms
among the matrix elements $\{\delta_{\alpha\beta}-\Sigma^{-1}_{\alpha\beta}\}$ 
are order $n$ and all the $n(n+1)$ out-of-diagonal terms are order $1$.
Then all terms in the exponent of eqs.(\ref{HA}),(\ref{HB}) are order $n$, 
and we can expand the
exponentials, which allows us to perform the quenched averages across 
$\{\varepsilon,\vartheta^0\}$.
Considering the expression of $\tilde{\eta}(\vartheta,s)$, eq.(\ref{tuning_tot}), 
one obtains:
\begin{eqnarray}
H^A(\tilde{z}_0^A,\tilde{z}_1^A)&\simeq& n
(\tilde{\tilde{z}}_0^A-\tilde{z}_1^A)\Lambda^1_\eta;\nonumber\\
H^B(\tilde{z}_0^B,\tilde{z}_1^B)&\simeq& n\left(
(\tilde{\tilde{z}}_0^B-\tilde{z}_1^B)\Lambda^1_\eta+\frac{\tilde{z}_1^B}{1+2\sigma^2\tilde{z}_1^B}\left[\Lambda_\eta^1-\Lambda^2_\eta\right]\right);
\label{HA_HB}
\end{eqnarray}
\begin{eqnarray}
\Lambda_\eta^1&=&\sum_s\int d\vartheta p(s,\vartheta)\langle[\tilde{\eta}(\vartheta,s)]^2\rangle_{\varepsilon,\vartheta^0}\nonumber\\
              &=&(\eta^0)^2\left[(A_2+\alpha^2-2\alpha A_1)\langle\varepsilon^2\rangle_{\varepsilon}+\alpha^2+2\alpha(A_1-\alpha)\langle\varepsilon\rangle_{\varepsilon}\right];
\label{Lambda1}
\end{eqnarray}
\begin{eqnarray}
\Lambda_\eta^2&=&\sum_{s_1,s_2}\int d\vartheta_1d\vartheta_2 [p(s,\vartheta)]^2 
\langle\tilde{\eta}(\vartheta_1,s_1)\tilde{\eta}(\vartheta_2,s_2)
\rangle_{\varepsilon,\vartheta^0}\nonumber\\
              &=&(\eta^0)^2\left[(A_1-\alpha)^2\left(\frac{K-1}{K}
\langle\varepsilon\rangle_{\varepsilon}^2+\frac{1}{K}\langle\varepsilon^2
\rangle_{\varepsilon}\right)+\alpha^2+2\alpha(A_1-\alpha)
\langle\varepsilon\rangle_{\varepsilon}\right];
\label{Lambda2}
\end{eqnarray}
\begin{equation}
A_1=\frac{1}{2^{2m}}\left(\begin{array}{c}2m\\m\end{array}\right);\,\,\,
A_2=\frac{1}{2^{4m}}\left(\begin{array}{c}4m\\2m\end{array}\right);\,\,\,
\alpha=\frac{\eta^f}{\eta^0}.
\label{A_2}
\end{equation}

A similar expansion in $n$ for $Tr\ln\Sigma(\tilde{z}_0,\tilde{z}_1)$ and for
$Tr\ln G(z_0,z_1)+F(z_0,z_1)$ allows to derive explicitely the saddle point equations:
\begin{eqnarray}
z_0^{A,B}&=&\sigma^2+\Lambda_\eta^1;\nonumber\\
z_1^A&=&\frac{2\sigma^4\tilde{z}_1^A}{2\sigma^2\tilde{z}_1^A+1}+\Lambda_\eta^1;\nonumber\\
z_1^B&=&\frac{2\sigma^4\tilde{z}_1^B}{2\sigma^2\tilde{z}_1^B+1}+
\left[1-\frac{1}{\left(1+2\sigma^2\tilde{z}_1^B\right)^2}\right]
\Lambda_\eta^1+\frac{1}{\left(1+2\sigma^2\tilde{z}_1^B\right)^2}\Lambda_\eta^2;\nonumber\\
\tilde{z}_1^{A,B}&=&-C\sigma^2_J\frac{r}{2}\left\{\sigma\left(\frac{\xi^0}{\sqrt{p+q}}\right)\frac{\xi^0}{\left(p+q\right)^{\frac{3}{2}}}-\frac{1}{p}\erf\left(\frac{\xi^0}{\sqrt{p+q}}\right)\right.\nonumber\\
&&\left.+\int_{-\infty}^{\infty} Dt
\left[1+\ln\left(\erf\left(-\frac{\xi^0-t\sqrt{q}}{\sqrt{p}}\right)\right)\right]\sigma\left(\frac{\xi^0-t\sqrt{q}}{\sqrt{p}}\right)\frac{1}{p^\frac{3}{2}}\left[\xi^0-t\frac{q+p}{\sqrt{q}}\right]\right \};\nonumber\\
\label{saddlepoint}
\end{eqnarray}
where:
$$
\erf(x)=\int_{-\infty}^{x} Dx^\prime=\int_{-\infty}^{x}dx^{\prime}\sigma(x^\prime);\,\,\,\,\,\,\,\sigma(x)=\frac{1}{\sqrt{2\pi}}e^{-\frac{x^2}{2}};
$$
$$
p=\sigma^2_\delta+C\sigma^2_J(z_0-z_1);\,\,\,\,\,q=C\sigma_J^2z_1.
$$

From the expression of $z_0^{A,B}$ in 
eq.(\ref{saddlepoint}), it is easy to verify that the dependence on 
$\tilde{\tilde{z}}_0^{A,B}$ in eq.(\ref{HA_HB}), which might affect the information in 
eq.(\ref{info_lim}), cancels
out with the products $z_0^A\tilde{\tilde{z}}_0^A$,$z_0^B\tilde{\tilde{z}}_0^B$ which 
should contribute to the information in the limit $n\rightarrow 0$ (see eq.(\ref{info_lim})).
Therefore, since $z_0^{A,B}$ is known and $z_1^{A,B}$ depends only on $\tilde{z}_1^{A,B}$, 
the mutual information can be expressed as a function of $z_1^{A,B}$,$\tilde{z}_1^{A,B}$, 
which in turn are to be determined self-consistently by the saddle point equations.

The average information per input cell can be written, finally:
\begin{equation}
\frac{I}{N}(\{\xi_i\},\vartheta\otimes
s)\!=\!\frac{1}{\ln 2}\left\{\tilde{z}_1^Bz_1^B-\tilde{z}_1^Az_1^A+r\left[\Gamma_1(z_0^B,z_1^B)\!-\!
\Gamma_1(z_0^A,z_1^A)\right]+
\Gamma_2^B(\tilde{z}_1^B)-\Gamma_2^A(\tilde{z}_1^A)\right\},
\label{final_info}
\end{equation}
with
\begin{eqnarray}
\Gamma_1(z_0,z_1)&=&-\sigma\left(\frac{\xi^0}{\sqrt{p+q}}\right)\frac{\xi^0p}
{2\left(p+q\right)^{\frac{3}{2}}}+
\frac{1}{2}\ln p\,\erf\left(\frac{\xi^0}{\sqrt{p+q}}\right)\nonumber\\
&-&\int_{-\infty}^{\infty} Dt\, \erf\left(-\frac{\xi^0-t\sqrt{q}}{\sqrt{p}}\right)\ln\left[\erf\left(-\frac{\xi^0-t\sqrt{q}}{\sqrt{p}}\right)\right];
\end{eqnarray}
\begin{equation}
\Gamma_2^A(\tilde{z}_1^A)=+\frac{1}{2}\ln (1+2\sigma^2 \tilde{z}_1^A)-\tilde{z}_1^A
(\sigma^2+\Lambda_\eta^1);
\end{equation}
\begin{equation}
\Gamma_2^B(\tilde{z}_1^B)=\frac{1}{2}\ln (1+2\sigma^2
\tilde{z}_1^B)-\tilde{z}_1^B (\sigma^2+\Lambda_\eta^1)+
\frac{\tilde{z}_1^B}{1+2\sigma^2\tilde{z}_1^B}\left[\Lambda_\eta^1-\Lambda_\eta^2\right].
\end{equation}

The expression for the mutual information only contains terms linear in either $N$ or $M$. 
Since the last of
the saddle-point equations, (\ref{saddlepoint}), contains $r$, if one fixes $N$ and increases $M$ the information
grows non-linearly, because the position of the saddle point varies. It turns out that, as shown below, the growth 
is only very weakly sublinear, at least when $M\le N$. Analogously, fixing $M$ and varying $N$ we would find a 
non-linearity due to the $r$-dependence of the saddle point. If $r$ is fixed and $N$ and $M$ grow together, 
the information rises purely linearly.

What our analytical treatment misses out, however, is the nonlinearity required to 
appear as the mutual information approaches its ceiling, the entropy of the stimulus set.
The approach to this saturating value was described at the input stage 
\cite{pap35,sompo+00}, where also the initial linear rise (in $N$) was obtained in 
the large noise limit \cite{vale+01b,pap35}. Our saddle point method is 
in same sense similar to taking a large (input) noise limit, $\sigma \to \infty$, to 
its leading (order $N/\sigma^2$) term. It is possible that the saddle point method 
could be extended, to account also for successive terms in a large noise expansion. 
This would probably require integrating out the fluctuations around the saddle point, 
but by carefully analysing the relation of different replicas to different values of 
the quenched variables. We leave this possible extension to future work. The present 
calculation, therefore, although employing a saddle point method which is usually 
applicable for large $N$ and $M$, should be considered effectively as yielding
the initial linear rise in the mutual information, the one observed with $M$ small.

\section{Numerical results}

Eq.(\ref{saddlepoint}) for $\tilde{z}_1^{A,B}$ has been solved numerically
using a Matlab Code. Convergence to self-consistency has been found already
after $50$ iterations with an error lower than $10^{-10}$.

\begin{figure}
\centerline{
\psfig{figure=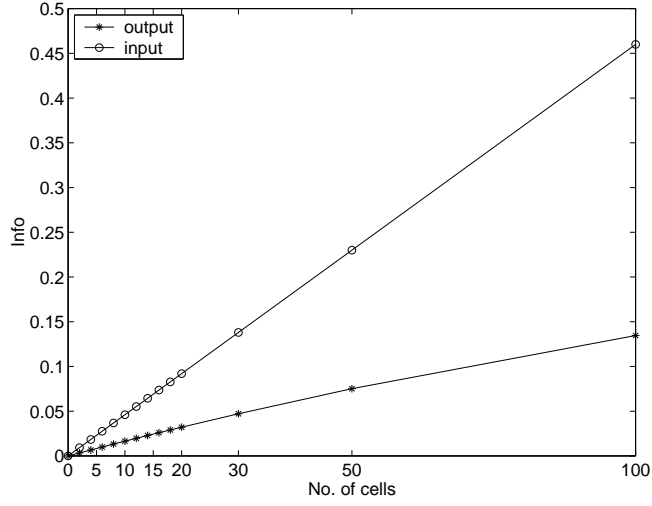,height=7cm,angle=0}}
\caption{Information rise, from eq.(\protect\ref{final_info}), as a function of
the number $M$ of output neurons. $N=100$; $K=4$; $(\eta^0)^2=0.1$; $\alpha=0.2$; 
$\xi^0=-0.4$; $\sigma^2=1$; $m=1$; $C\sigma_J^2=1$; $\sigma^2_\delta=1$. The 
distribution $\varrho(\varepsilon)$ in eq.(\protect\ref{ro_eps}) is just equal to 
$1/3$ for each of the $3$ allowed $\varepsilon$ values of $0$, $1/2$ and $1$. The 
upper curve is the linear term in the input information,
calculated as a function of $N$ as in \cite{vale+01b} with identical parameters.} 
\label{fig1} 
\end{figure}

Fig.\ref{fig1} shows the mutual information as a function of the output population size, 
for an input population size equal to $100$ cells. This is contrasted with the information 
in the input units, about exactly the same set of correlates, calculated as in \cite{vale+01b}, 
by keeping only the leading (linear) term in $N$. In fact, in \cite{vale+01b} the mutual
information carried by a finite population of neurons firing according to eq.(\ref{dist})
had been evaluated analytically, in the limit of large noise, by means of an expansion 
in $N(\eta^0)^2/4\sigma^2$. To linear order in $N$ the analytical expression for the 
information carried by $N$ input neurons reads:  
\begin{equation} 
I_{input}(\{\eta_j\},\vartheta\otimes s)=\frac{1}{\ln 2}\frac{N}{2\sigma^2}\left(\Lambda_\eta^1-\Lambda_\eta^2\right); 
\label{input} 
\end{equation} 
where $\Lambda_\eta^{1}$, $\Lambda_\eta^{2}$ are defined, again, as in eqs.(\ref{Lambda1}),
(\ref{Lambda2}). In analogy to what had been done in \cite{Tre+95} we have
set  $C\sigma^2_J=1$. As evident from the graph, also the output information
is essentially  linear up to a value of $r\simeq 0.5$, and quasi-linear even
for $r=1$. It should be  remined, again, that our saddle point method only
takes into account the term linear  in $N$ in the information {\em input}
units carry about the stimulus. It is not possible, therefore, for
eq.(\protect\ref{final_info}) to reproduce the saturation in the mutual 
information as it approaches the entropy of the stimulus set (which is finite,
if one  considers only discrete stimuli). The nearly linear behaviour in $M$
thus reflects the  linear behaviour in $N$ induced, in the intermediate
quantity (the information available at the input stage), by our saddle point
approximate evaluation.

As it is clear from the comparison in Fig.\ref{fig1}, when the two populations of 
units are affected by the same noise the input information is considerably higher 
than the output one. This is expected, since output and input noise sum up while 
influencing the firing of output neurons, but also because the input distribution 
is taken to be a pure gaussian, while the output rates are affected
by a threshold. If the input-output tranformation were linear and the output noise 
much smaller than the input one, one would expect that output and input units would 
carry the same amount of information.

Briefly, in a linear network with zero output noise one has:
\begin{equation}
p(\{\xi_i\}|\{\eta_j\})=\prod_i\delta(\xi_i-\sum_jc_{ij}J_{ij}\eta_j);
\end{equation}

Considering eqs.(\ref{xi_eta}),(\ref{dist}), an {\it effective} expression for the 
distribution 
$p(\{\xi_i\}|\vartheta,s)$ can be obtained by direct integration of the $\delta$ 
functions $\delta(\xi_i-\sum_jc_{ij}J_{ij}\eta_j)$ via their integral representation, on $\{\eta_j\}$:
\begin{equation}
p(\{\xi_i\}|\vartheta,s)=\frac{1}{\sqrt{(2\pi)^Mdet\Sigma}}e^{-\sum_{i,j}(\xi_i
-\tilde{\xi}_i(\vartheta,s))(\Sigma^{-1}_{ij})/2(\xi_j
-\tilde{\xi}_j(\vartheta,s))};
\end{equation}
\begin{equation}
\tilde{\xi}_i(\vartheta,s)=\sum_jc_{ij}J_{ij}\tilde{\eta}_j(\vartheta,s);
\end{equation}
\begin{equation}
\Sigma_{ij}=\sigma^2\sum_k c_{ik}J_{ik}c^T_{kj}J^T_{kj};
\end{equation}

This distribution is then used to evaluate both the equivocation, eq.(\ref{equiv}), and 
the entropy of the responses, eq.(\ref{outent}). We do not report the calculation, that 
is straightforward and analogous to the one reported in \cite{vale+01b}. The final result, 
which is valid for a finite population size $M$, and up to the linear approximation in 
$M(\eta^0)^2/4\sigma^2$, is analogous to eq.(\ref{input}):
\begin{equation}
I_{lin}(\{\xi_i\},\vartheta\otimes s)=\frac{1}{\ln 2}\frac{M}{2\sigma^2}\left(\Lambda_\eta^1-\Lambda_\eta^2\right);
\label{asymptot}
\end{equation}
Thus, we expect that taking the limits $\xi^0\rightarrow\infty$ and $r\rightarrow 0$ 
simultaneously in eq.(\ref{final_info}), we should get to the same result: the output 
information should equal the input one when $\sigma^2$ grows large.

\begin{figure}
\centerline{
\psfig{figure=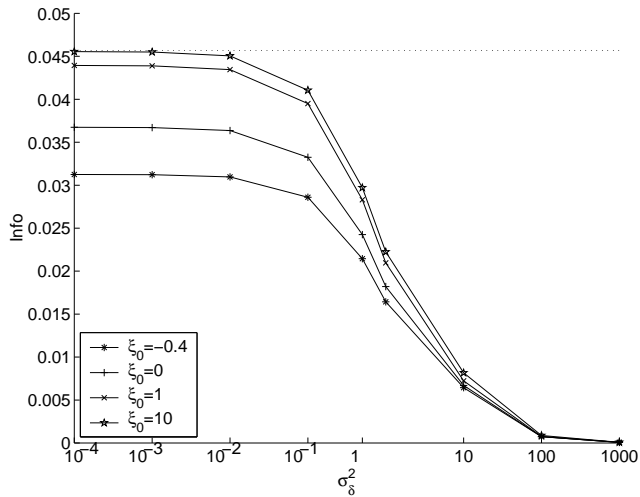,height=7cm,angle=0}}
\caption{Output information, from eq.(\protect\ref{final_info}), as a function of
the output noise $\sigma^2_\delta$, for 4 different values of the output (reciprocal) 
threshold $\xi^0$. Logarithmic scale. $N=100$; $K=4$; $M=10$; $(\eta^0)^2=0.1$; 
$\alpha=0.2$; $\sigma^2=1$; m=1; $C\sigma_J^2=1$. The distribution 
$\varrho(\varepsilon)$ in eq.(\protect\ref{ro_eps}) is just equal to $1/3$ for each of 
the $3$ allowed $\varepsilon$ values of $0$, $1/2$ and $1$. The dotted line represents 
the asymptotic value of the input information, eq.(\ref{input}), for $N=10$.} 
\label{fig3} 
\end{figure}

From eq.(\ref{final_info}) it is easy to show that:
\begin{equation}
\lim_{r\rightarrow 0}\lim_{\xi^0\rightarrow\infty}I(\{\xi_i\},\vartheta\otimes
s)=\frac{1}{\ln 2}\frac{M}{2}\ln\left[1+\frac{2C\sigma^2_J\left(\Lambda_\eta^1-
\Lambda_\eta^2\right)} {\sigma^2_\delta +2C\sigma^2_J\sigma^2}\right];
\end{equation}
When $\sigma^2\gg\sigma^2_\delta,\Lambda_\eta^1,\Lambda_\eta^2 $ one obtains exactly 
the linear limit, eq.(\ref{asymptot}). We have verified this analytical limit by studying 
numerically  the approach to the asymptotic value of the mutual information. 
Fig.\ref{fig3} shows the dependence of output information on the output noise 
$\sigma_\delta^2$, for 4 different choices of the (reciprocal of the) threshold, 
$\xi^0$. A large value, $\xi^0=10$, implies linear output units. As expected, the 
output information, which always grows for decreasing values of the output noise, 
for $\xi^0=10$ approaches asymptotically the input information. For increasing values
of the output noise, the information vanishes with a typical sigmoid curve, with
its point of inflection when the output matches the input noise.

\begin{figure}
\centerline{
\psfig{figure=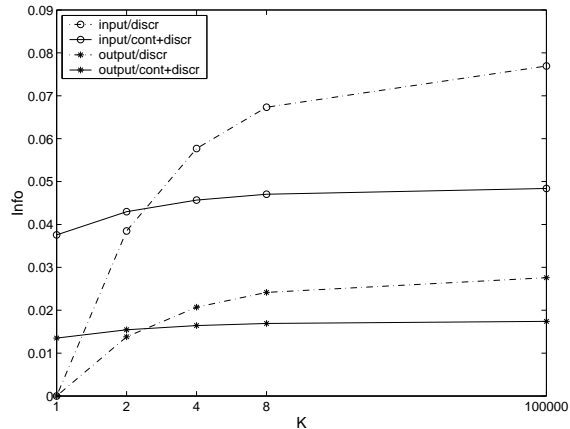,height=6cm,angle=0}}
\caption{Comparison between input and output information as a function of the number $K$
of discrete correlates, for the case of continuous+discrete correlates ($m=1$) or with 
purely discrete correlates (obtained by setting $m=0$). In eq.(\protect\ref{final_info})
we have set $N=100$; $r=0.1$; $\xi^0=-0.4$; $(\eta^0)^2=0.1$; $\alpha=0.2$;
$\sigma^2=1$; $C\sigma_J^2=1$; $\sigma^2_\delta=1$. The distribution
$\varrho(\varepsilon)$ in eq.(\protect\ref{ro_eps}) is just equal to $1/3$ for
each of the $3$ allowed $\varepsilon$ values of $0$, $1/2$ and $1$.} 
\label{fig4}  
\end{figure}

We have then examined how the information in output (compared to the input) depends 
on the number $K$ of discrete correlates and on the width of the tuning function 
(\ref{tuning2}), parametrized by $m$, with respect to the continuous correlate.
Fig.\ref{fig4} shows a comparison between input and output information for a sample 
of 10 cells, as a function of $K$. Both curves quickly reach an asymptotic value, 
obtained by setting $K\to\infty$ in eq.(\protect\ref{Lambda2}) for $\Lambda^2_{\eta}$.
The relative information loss in output is roughly constant with $K$. A comparison is 
shown with the case where correlates are purely discrete, which is obtained by setting 
$m=0$ in eq.(\ref{tuning2}). The curves exhibit a similar behaviour, even if the rise 
with $K$ is steeper, and the asymptotic values are higher. This may be surprising, but
it is in fact a consequence of the specific model we have considered, 
eq.(\protect\ref{tuning_tot}), where a unit has the same tuning curve to each of
the discrete correlates, only varying its amplitude with respect to a value constant
in the angle. As $K\to\infty$, most of the mutual information is about the
discrete correlates, and the tuning to the continuous dimension, present for $m=1$,
effectively adds noise to the discrimination among discrete cases, noise which is not
present for $m=0$. 

\begin{figure}
\centerline{ \psfig{figure=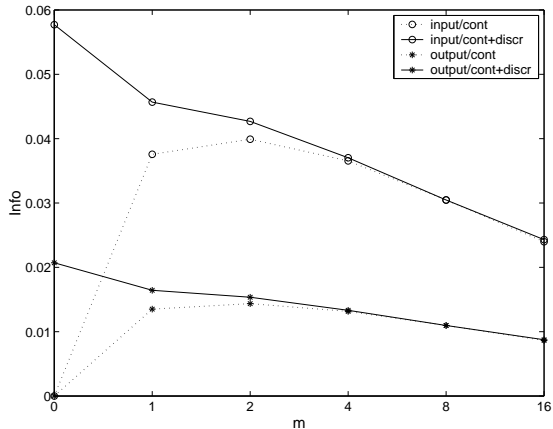,height=6cm,angle=0}}
\caption{Comparison between input and output information as a function of the 
selectivity along the continuous dimension, which is made sharper by increasing $m$. 
$K=1$ implies a purely continuous correlate, while the continuous+discrete case is 
obtained by setting $K=4$. In eq.(\protect\ref{final_info}) we have set $N=100$; 
$r=0.1$; $\xi^0=-0.4$; $(\eta^0)^2=0.1$; $\alpha=0.2$; $m=1$; $C\sigma_J^2=1$;
$\sigma^2_\delta=\sigma^2=1$, in both cases. The distribution
$\varrho(\varepsilon)$ in eq.(\protect\ref{ro_eps}) is just equal to $1/3$ for
each of the $3$ allowed $\varepsilon$ values of $0$, $1/2$ and $1$.} 
\label{fig5}    
\end{figure}

With respect to the continuous dimension, the selectivity of the input units can be 
increased by varying the power $m$ of the cosine from 0 (no selectivity) through 1 
(very distributed encoding, as for the discrete correlates) to higher values 
(progressively narrower tuning functions). Fig.\ref{fig5} reports the resulting
behaviour of the information in input and in output, for the case $K=1$ 
(only a continuous correlate) and $K=4$ (continuous+discrete correlates). Increasing
selectivity implies a "sparser" \cite{Tre+90} representation of the angle, the 
continuous variable, and hence less information, on average. However if the correlate
is purely continuous there is an initial increase, before reaching the optimal
sparseness. It should be kept in mind, again, that the asymptotic equality of the $K=1$
and $K=4$ cases is a consequence of the specific model, eq.(\protect\ref{tuning_tot}), 
which assigns the same preferred angle to each discrete correlate. The resolution with
which the continuous dimension can be discriminated does not, within this model, 
improve with larger $K$, while the added contribution, of being able to discriminate
among discrete correlates, decreases in relative importance as the tuning becomes sharper.

Figures \ref{fig4} and \ref{fig5} show that, as long as the output noise is non zero 
and the threshold is finite, information is lost going from input to output, but
the information loss does not appear to depend on the structure and on the 
dimensionality of the correlate.

Note that, while the purely continuous case has been easily obtained by setting $K=1$ 
in the expression of $\Lambda_\eta^2$, eq.(\ref{Lambda2}), for the purely discrete 
case it is enough to set $m=0$.

\section{Simulation results: information estimates via decoding }
One way to test the range of validity of our analytical approximation si via 
numerical estimates of the information.
In our specific case where we deal with populations of continuous neurons and with 
$4$ different sources of quenched disorder, direct numerical integration is prohibitive.
A more feasible method is to generate simulated data both in input and in output 
and then to estimate the information from the data, resorting to some algorithm.

Many previous studies (revised in \cite{Rol+98}) have shown that information 
estimates from data are extremely sensitive to sampling and the distortion is 
the more serious the larger is the space of the possible configurations which has 
to be sampled. In our case where neurons are characterized by their 
time averaged continuous firing rates, a simple binning of the responses into 
discrete distributions results in a considerable distortion, which becomes even more 
serious since a sparse sampling is required to perform the average across the four 
quenched distributions.

Several procedures have been proposed to correct the bias affecting information estimates; 
some of them (see \cite{Rol+98}) are based on regularizations like binning or smoothing; 
other ones rely on more theoretical approaches aiming at providing an analytical 
expression for the correction \cite{Pan+96b}.

When the responses vary in a continuum and the population size is very large 
it has been suggested that the best estimate is obtained via {\em decoding}:
the method consists in generating a {\em predicted }stimulus $s^\prime(\bo r_k)$
from each simulated response vector $\bo r_k(s)$ in each trial $k$ by matching 
$\bo r_k(s)$ to the average response vector $\bo r_{av}(s^\prime)$, for all stimuli.
The predicted stimulus will be the one corresponding to the best match. 
Summing on all the trials one can derive a probability table $p(s,s^\prime)$; then,
the mutual information between the true and the pseudo-stimuli is computed instead of the original one, between
stimuli and responses:

\begin{equation}
I(s,s^\prime)=\sum_{s=1}^p\sum_{s^\prime=1}^p P(s,s^\prime)\log_2
\frac{P(s,s^\prime)}{P(s)P(s^\prime)}.
\label{info_dec}
\end{equation}

From a theoretical point of view the optimal transformation to derive 
the probability $p(s^\prime|\bo r)$ is defined by Bayes rule (Bayesian decoding):

\begin{equation}
p(\bo r|s) p(s)=p(\bo r)p(s|\bo r);
\label{bayes}
\end{equation}

In our specific case the input distribution is defined in eq.(\ref{dist}), 
so that relationship (\ref{bayes}) can be explicitely inverted; 
on the contrary the output distribution is not explicitly known and
one has to fit some function to the responses to be able to invert the relationship 
(\ref{bayes}). Further details about these procedures can be found in \cite{Rol+98,Rol+97a}.

\begin{figure}
\centerline{ \psfig{figure=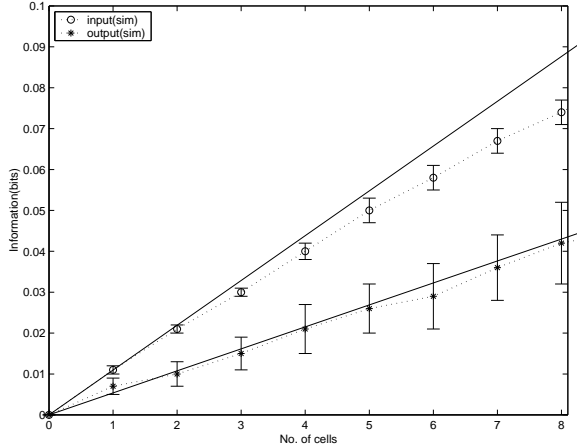,height=6cm,angle=0}}
\caption{Comparison between the analytical approximation (solid lines) 
and simulations (dashed lines) for the input and the output information. 
$N=1000$; $K=4$; $r=0.02$; $\xi^0=2$; $(\eta^0)^2=0.1$; $\alpha=0.1$; $m=0$; 
$C\sigma_J^2=1$; $\sigma^2_\delta=\sigma^2=1$.The distribution
$\varrho(\varepsilon)$ in eq.(\protect\ref{ro_eps}) is just equal to $1/2$ for
each of the $2$ allowed $\varepsilon$ values of $0$ and $1$ } 
\label{fig6}    
\end{figure}

Fig.\ref{fig6} shows the results of simulations for a sample of $20$ output cells 
receiving from $1000$ input cells.
A MATLAB code has been used to generate data in an amount of $2000$ responses per neuron per stimulus, 
where we have considered the simpler case of $4$ purely discrete stimuli ($m=0$ in eq.\ref{tuning2}).
The decoding in output has been optimized choosing a high value for the threshold $\xi^0$, 
so that the single cell output distribution, as derived from the simulated data, 
could be roughly fitted by a gaussian. 
For each population size we have chosen many random subsamples of units out of the 
whole population, both in input and in output, and we have then averaged the information 
value across subsamples. Quenched averages have been performed resorting to a sparse sampling.

The plot shows that the curves obtained via simulations (dashed lines) match 
the analytical prediction (solid lines) for a very small number of cells, 
but they deviate when the number of cells increases; since both the input and 
the output noise are large and the information values are much lower than the upper bound 
of $2$ bits, this mismatch cannot be due to a deviation from linearity close to the ceiling regime. 
It is more probably an effect due to the distortion caused by the 
decoding, which is known to increase with the population size. 
The discrepancy between the true and the decoded information has also been recently 
investigated and quantified analytically \cite{ines+01b}.
 
The analytical approximation seems to have a wider applicability in output, 
where even for a population size larger than $4$-$5$ cells the analytical curve 
falls within the error. This is mainly an effect due to the larger error 
characterizing the output information, which in turn is due to the stronger 
fluctuations in the information values across the different subsamples of cells.  
We have checked that the analytical results are always confirmed for a population 
size of $1$ or $2$ cells varying the number of discrete correlates and the value of 
the input and output noise. 

We certainly cannot conclude from this that our analytical 
approximation is not valid for population sizes larger than $2$-$3$ cells.
On the contrary we are currently devising new algorithms to reduce the bias in the 
information estimates and preliminary results seem to suggest a much wider range of 
 validity of the analytical 
approximation. A more detailed study comparing decoding to other algorithms will be published elsewhere \cite{Isaac+02b}.

\section{Discussion}
We have attempted to clarify how information about multi-dimensional
stimuli, with both a continuous and a discrete dimension, is transmitted from a 
population of units with a known coding scheme, down to the next stage of processing.

Previous studies had focused on the mutual information between input and output units 
in a two-layer threshold-linear network either with learning \cite{Tre+95} or
with simple  random connection weights \cite{Tre+98b}.

More recent investigations have tried to quantify the efficiency of a population of 
units in coding a set of discrete \cite{pap35} or continuous \cite{sompo+00} correlates.
The analysis in \cite{pap35} has been then generalized to the more realistic case of 
multi-dimensional continuous+discrete correlates \cite{vale+01b}. 

This work correlates with both research streams, in an effort to define a unique 
conceptual framework for population coding. The main difference with the second group 
of studies is obviously the presence of the network linking input to output units. The 
main difference with the first two papers, instead, is the analysis of a distinct 
mutual information quantity: not between input and output units, but between correlates
("stimuli") and output units. In \cite{pap35} it had been argued, for a number $K$ of 
purely discrete correlates, that the information {\em about} the stimuli reduces
to the information about the "reference" neural activity when $K\to\infty$. 
The reference activity is simply the mean response to a given stimulus when the 
information is measured from the variable, noisy responses around that means; or it 
can be taken to be the stored pattern of activity, when the retrieval of such patterns 
is considered, as in \cite{Tre+95}. True, the information about the stimuli
saturates at the entropy of the stimulus set, but for $K\to\infty$ this
entropy diverges, only the  linear term in $N$ is relevant \cite{pap35}, and
the two quantities, information about  the stimuli and information about the
reference activity, coincide.

Our present saddle point calculation is only able to capture, effectively, the mutual 
information which is linear in the number of input units, as mentioned above. It fails 
to describe the approach to the saturating value, the entropy of the set of correlates, 
be this finite or infinite. Therefore, ours is close to a calculation of the 
information about a reference activity - in our case, the activity of the input units. 
The remaining difference is that we can take into account, albeit solely in the linear
term, the dependence on $K$ (through the equation for $\Lambda^2_{\eta}$, 
eq.(\ref{Lambda2})), without having to take the further limit $K\to\infty$. 

Due to the presence of a threshold and of a non zero output noise the information in 
output is lower than that in input, and we have shown analytically that in the limit 
of a noiseless, linear input-output transfer function the ouptput information tends 
asymptotically to the input one. We have not, however, introduced a threshold in the 
input units, which would be necessary for a fair comparison. In an independent line 
of research, recent work \cite{vale+01a} has also quantified the contribution to the 
mutual information, in a different model, of cubic and higher order non-linearities 
in the transfer function, by means of a diagrammatic expansion in a noise parameter.
In \cite{vale+01b} it has been shown that the effect of a threshold in the input units 
on the input information results merely in a renormalization of the noise. The resulting 
effect on the output information remains to be explored, possibly with similar methods.

Considering mixed continuous and discrete dimensions in our stimulus set, we had been 
wondering whether the information loss in output depended on the presence or absence 
of discrete or continuous dimensions in the stimulus structure. We have shown that 
for a fixed, finite level of noise this loss does not depend significantly on the 
structure of the stimulus, but solely on the relative magnitude of input and output 
noise, and on the position of the output threshold.

Our analytical efforts have been also motivated by the difficulty to perform a 
simulation study for this specific scheme of continuous rate coding in presence of several 
distinct sources of quenched disorder. 
Nonetheless we have performed some simulations using a {\it decoding} procedure to 
estimate the information from simulated data both in input and in output. Our results 
confirm previous findings in that the distortion due to decoding grows with the 
population size, so that the simulations confirm the anlytical prediction only for a 
very small number of cells. 
We are currently devising new computational methods to improve the match between the 
analytical and the simulation results.

Further developments of this analysis include the evaluation of the output information
in presence of learning, in line with \cite{Tre+95}, and with correlations in
the firing  of input units.

A recent work has shown that the interplay between short and long range
connectivities in the Hopfield model leads to a deformation of the phase
diagram with the appearence of novel phases \cite{Ska+01}.
It would be interesting to introduce short and long range connections in our model,
and to examine how the coding efficiency of output
neurons depends on the interaction between short and long range connections.
This will be the object of future investigations.

\subsection*{Acknowledgements}
We have enjoyed extensive discussions with In\'es Samengo and Elka Korutcheva. 
Valeria Del Prete thanks Isaac Perez-Castillo for very useful comments.
Partial support from Human Frontier Science Programme grant RG 0110/1998-B.


\end{document}